\documentclass[12pt,a4paper]{article}

\usepackage[margin=2.5cm]{geometry}
\usepackage{times}
\usepackage{amsmath,amssymb,amsthm}
\DeclareMathOperator{\sgn}{sgn}
\usepackage{graphicx,booktabs,tabularx}
\usepackage{algorithm}
\usepackage{algpseudocode}
\usepackage[numbers]{natbib}
\usepackage{url}
\usepackage{xcolor}
\usepackage{physics}
\usepackage[left]{lineno} 

\usepackage{float}
\usepackage{placeins} 
\usepackage{yhmath}
\usepackage{subfigure}
\usepackage{mathdots}
\usepackage{MnSymbol}
\usepackage{dsfont}
\usepackage{soul}
\usepackage[normalem]{ulem}
\usepackage{hyperref}
\hypersetup{
  hidelinks,
  pdftitle={Exact Spin Elimination for Quadratic and k-Local Ising Optimization},
  pdfauthor={Natalia G. Berloff}
}
\usepackage{bm}
\usepackage{multirow}

\usepackage{mathtools}
\usepackage{microtype}

\graphicspath{{figures/}{figure_revision/figures/}{figs_v38/}{figs/}{./}}

\newcommand{\widefiguregraphic}[1]{%
  \makebox[\textwidth][c]{\includegraphics[width=180mm]{#1}}%
}

\newtheorem{proposition}{Proposition}[section]
\newtheorem{corollary}[proposition]{Corollary}

\newcommand{\Erefmain}{E_{\mathrm{ref}}^{\mathrm{main}}}

\newcommand{\Erefho}{E_{\mathrm{ref}}^{\mathrm{HO}}}
\newcommand{\TTSindex}{\mathrm{TTS}_{90}^{\mathrm{pool}}}

\makeatletter
\newcommand{\beginsupplement}{%
  \clearpage
  \setcounter{page}{1}%
  \setcounter{section}{0}%
  \renewcommand{\thesection}{\arabic{section}}%
  \renewcommand{\@seccntformat}[1]{\csname SIprefix##1\endcsname\csname the##1\endcsname\quad}%
  \setcounter{figure}{0}\renewcommand{\thefigure}{\arabic{figure}}%
  \setcounter{table}{0}\renewcommand{\thetable}{\arabic{table}}%
  \setcounter{equation}{0}\renewcommand{\theequation}{S\arabic{equation}}%
  \renewcommand{\figurename}{Supplementary Figure}%
  \renewcommand{\tablename}{Supplementary Table}%
  \renewcommand{\theHsection}{S\arabic{section}}%
  \renewcommand{\theHfigure}{S\arabic{figure}}%
  \renewcommand{\theHtable}{S\arabic{table}}%
  \renewcommand{\theHequation}{S\arabic{equation}}%
  \setcounter{proposition}{0}\renewcommand{\theproposition}{S\arabic{proposition}}%
  \renewcommand{\theHproposition}{S\arabic{section}.\arabic{proposition}}%
  \setcounter{algorithm}{0}\renewcommand{\thealgorithm}{S\arabic{algorithm}}%
  \floatname{algorithm}{Supplementary Algorithm}%
}

\makeatother
\begin{document}


\title{Exact Spin Elimination for Quadratic and $k$-Local Ising Optimization}

\author{Natalia G. Berloff$^{1,*}$}
\date{}
\maketitle

\begin{abstract}
Ising solvers have a finite spin budget. Quadratization uses auxiliary
spins to replace higher-order interactions by pairwise ones. We show
that removing spins while allowing more complex interactions can fit
larger problems within the same spin budget and improve optimization. Walsh
elimination minimizes over one spin at a time, writes the resulting
interaction as a sum of spin products in the Walsh basis, and stores
a rule for recovering the removed spin. Explicit resource
limits control which eliminations are accepted. In a held-out test with the protocol frozen before instance generation and one annealing schedule shared by both formulations, mean observed success rises from 10.4\% to 97.5\% on sparse quadratic spin glasses and from 16.9\% to 87.5\% on three-body problems. Observed success increases on every instance with a verified reference minimum, and pooled time-to-solution estimates decrease by factors of about 34 and 11, respectively, with preprocessing included. For zero-field pairwise models on connected simple cubic graphs with more than four vertices, we prove that a fixed spin budget accommodates 50\% larger inputs while retaining pairwise interactions, zero fields, residual degree at most six, and no increase in coupling count. These results establish exact spin elimination as a controlled trade between active-spin capacity and interaction complexity.
\end{abstract}

\noindent $^{1}$Department of Applied Mathematics and Theoretical Physics,
University of Cambridge,\\
Wilberforce Road, Cambridge CB3 0WA, United Kingdom\\
\noindent $^{*}$email: \textit{N.G.Berloff@damtp.cam.ac.uk}

\section{Introduction}
\label{sec:introduction}

Many discrete optimization problems can be expressed using Ising
spins $s_i\in\{-1,+1\}$~\cite{boros2002pseudo,mohseni2022ising,Neukart2017}.
Let $\mathbf{s}=(s_1,\ldots,s_N)$ be a configuration of $N$ spins and
$[N]=\{1,\ldots,N\}$. A general $k$-local Hamiltonian is
\begin{equation}
\label{eq:ising-general-intro}
H(\mathbf{s})
=
\sum_{\substack{S\subseteq[N]\\ |S|\leq k}}
c_S\prod_{i\in S}s_i,
\end{equation}
where $c_S\in\mathbb{R}$ is the coefficient of the term on $S$, and $k$
is the largest number of spins in one term. The case $k=2$ is the pairwise
Ising model. The task is to find
\begin{equation}
\label{eq:ising-ground-intro}
E_0=\min_{\mathbf{s}\in\{-1,+1\}^{N}}H(\mathbf{s}).
\end{equation}
A configuration attaining $E_0$ is a ground state. Finding a ground state
is NP-hard in general~\cite{Barahona1982}.

The cost of solving an Ising problem depends on both the number of active
spins and the structure of their interactions
\cite{DwaveMapping,egger2021,aramon2019physics,tatsumura2021scaling,
Inagaki2016,Mcmahon2016,Pierangeli2019,camsari2017stochastic,
mohseni2022ising}. When a solver lacks the required connections, mapping
the problem onto its available connections can require many extra
spins~\cite{konz2021embedding}. Quadratization replaces higher-order
terms by pairwise terms and auxiliary spins
\cite{rosenberg1975reduction,valiante2020scaling}. Platforms that implement
multi-spin interactions directly avoid these auxiliary spins but still
have limits on spin count and interaction order
\cite{stroev2021discrete,hizzani2024memristor,
bhattacharya2024computing,pedersen2019native,andrade2022engineering,
katz2023demonstration}. We ask whether spins can instead be removed exactly
before optimization without making the interactions they create too costly.

Removing spins can make a larger problem fit within a fixed spin limit.
To improve search as well, the reduction must save more work than it
adds. Preparing the reduced model (preprocessing) takes time, and
elimination can add terms, increase the number of spins in each term,
and widen the range of coefficient values. Elimination preserves the
minimum energy in exact arithmetic, but does not guarantee faster search.

For a selected spin $s_a$, write
\begin{equation}
\label{eq:intro-local-block}
H(\mathbf{s})
=
H_{\setminus a}(\mathbf{s}_{\setminus a})
+s_aP_a(\mathbf{s}_{\mathcal{N}_a}).
\end{equation}
Here $\mathbf{s}_{\setminus a}$ contains all spins other than $s_a$,
and $H_{\setminus a}$ contains all terms without $s_a$. The local
polynomial $P_a$ depends only on $\mathbf{s}_{\mathcal{N}_a}$, where
the neighbourhood $\mathcal{N}_a$ consists of the spins that share a
term with $s_a$. For a fixed configuration of these neighbouring spins,
\begin{equation}
\label{eq:intro-local-minimum}
\min_{s_a\in\{-1,+1\}}s_aP_a
=-|P_a|.
\end{equation}
Walsh elimination replaces the terms $s_aP_a$ that contain the selected
spin by $-|P_a|$. It writes this function as a sum of products of
neighbouring spins, the Walsh basis, and stores the recovery rule
$s_a=-\sgn P_a$, with both values allowed when $P_a=0$. Repeating the
step gives the minimum energy over the removed spins for each
configuration of the remaining spins. Applying the stored rules in
reverse order recovers an original configuration of equal energy from
every reduced configuration.

Eliminating a spin with $d_a$ neighbours requires a table of $2^{d_a}$
values and can create up to $2^{d_a}$ coefficients. The compiler therefore
limits the number of neighbours of a spin selected for removal, the
order of the new terms, and the growth in their number. It reports
how many spins were removed and how long preprocessing took. This
reverses the trade made by quadratization: fewer variables can require
higher-order terms.

Roof duality and reductions based on equality constraints fix variables
only when a certificate or constraint is available
\cite{boros2002pseudo,hahn2017reducing,shirai2023spin}. By contrast,
sample-persistence variable reduction (SPVAR) and methods that use
sampled correlations to reduce the problem recursively choose values
from samples. One incorrect fixing decision can then exclude every
optimum~\cite{karimi2017effective,bravyi2020obstacles,
finvzgar2024quantum}. Renormalization and decimation are generally
approximate~\cite{schollwock2005density,refael2002spin}.

Exact elimination and recovery of variables are established in
pseudo-Boolean optimization~\cite{Hammer1969,Crama1990}, as reviewed by
Boros and Hammer~\cite[Section~4.6]{boros2002pseudo}. Sparse spin-glass
methods use exact reductions that retain pairwise
interactions~\cite{boettcher2003numerical,boettcher2008reduction,boettcher2024physics}.
The Appendix of Boettcher and Davidheiser~\cite{boettcher2008reduction}
also gives the general elimination rule and its Walsh coefficients.
The minimization in Eq.~\eqref{eq:intro-local-minimum} is also the
elementary step of min-sum bucket elimination
\cite{bertele1972nonserial,dechter1999bucket}. Eliminating every variable
makes this an exact solver, but its cost grows exponentially with the
largest number of remaining neighbours encountered at any step, including
connections created by earlier eliminations. This number is the induced
width $w$. Choosing the best order does not always keep this cost small:
for sparse random expanders, even the smallest possible $w$ (the graph's
treewidth) grows in proportion to the number of
spins~\cite{bollobas1988isoperimetric,grohe2009tree}. Mini-bucket elimination
bounds the width by approximating the eliminated function
\cite{dechter2003minibuckets}. Our approach retains the exact function,
stops when it reaches the chosen resource limits, and passes the remaining
problem to a solver that accepts higher-order terms.

We address three questions: does elimination improve search when its cost
is included, do the gains persist on new instances, and can a reduction in
spin count be guaranteed while retaining pairwise interactions?

Development tests on sparse quadratic and higher-order problems with
certified minima show that eliminating spins can improve search even
when preprocessing is included. Tests with parallel tempering and a
second solver implementation also favour elimination. We then fix and
record the protocol before generating new instances, and use the same
annealing schedule for both formulations. This test finds a higher
fraction of successful attempts on all $39$ quadratic and $40$
three-body instances with verified minima. The test starts with $40$
instances of each type; verifying the minimum of one quadratic instance
exceeds the time limit (Section~\ref{sec:heldout}).
In a separate comparison with roof duality and SPVAR that does not require
a target energy, elimination removes $30$--$40\%$ of the spins and
returns an energy no worse than the unreduced baseline on any of $60$
instances under each tested protocol. For pairwise models with no local
fields, we also prove that a fixed spin budget can accommodate $50\%$
larger inputs. This result applies to connected simple graphs with
more than four vertices in which every vertex has three neighbours
(cubic graphs). The reduced model remains pairwise and has no local
fields. Each spin has at most six neighbours, and the number of
couplings does not increase. The solver must support the required
couplings. These gains have limits: preprocessing and the additional
interactions created by elimination can outweigh the benefit of removing
spins.

The Results establish the elimination and recovery rules, describe the
resource limits, and test their effect on search and spin capacity.
The Methods explain the numerical settings and comparisons. The
Supplementary Information \cite{si} gives full protocols, comparisons with related
methods, worked examples, and additional checks and statistics.

\section{Results}
\label{sec:results}

\subsection{Exact Walsh elimination}
\label{sec:klocal-theory}

\emph{One exact elimination step.} In the construction introduced in
Eqs.~\eqref{eq:intro-local-block}--\eqref{eq:intro-local-minimum},
the neighbourhood $\mathcal{N}_a$ of a spin $s_a$ consists of all spins
that occur with it in at least one term with $c_S\neq0$, and its local
polynomial is
\begin{equation}
\label{eq:local-block}
P_a(\mathbf{s}_{\mathcal{N}_a})
=
\sum_{\substack{S\subseteq[N]\\a\in S}}
c_S
\prod_{i\in S\setminus\{a\}}s_i .
\end{equation}
The empty product for $S=\{a\}$ equals one, so the definition includes
a local field, a term linear in $s_a$. With $\mathbf{s}_{\setminus a}$
denoting all other spins
and $H_{\setminus a}(\mathbf{s}_{\setminus a})$ the sum of terms that
do not contain $s_a$, we have $H=H_{\setminus a}+s_aP_a$, as in
Eq.~\eqref{eq:intro-local-block}. For each fixed assignment of
$\mathbf{s}_{\setminus a}$, Eq.~\eqref{eq:intro-local-minimum} then
gives the reduced Hamiltonian
\begin{equation}
\label{eq:reduced-H}
H^{\setminus a}(\mathbf{s}_{\setminus a})
\equiv
\min_{s_a\in\{-1,+1\}}H(\mathbf{s})
=
H_{\setminus a}(\mathbf{s}_{\setminus a})
-
\left|P_a(\mathbf{s}_{\mathcal{N}_a})\right| .
\end{equation}
The subscript in $H_{\setminus a}$ means that the terms containing
$s_a$ are omitted; the superscript in $H^{\setminus a}$ means that
those terms are replaced by their minimum over $s_a$.
The eliminated spin is recovered from
\begin{equation}
\label{eq:backsub}
s_a^\star
=
\begin{cases}
-\sgn\!\left[P_a(\mathbf{s}_{\mathcal{N}_a})\right],
& P_a(\mathbf{s}_{\mathcal{N}_a})\neq 0,
\\
-1 \text{ or } +1,
& P_a(\mathbf{s}_{\mathcal{N}_a})=0.
\end{cases}
\end{equation}
At a tie, both signs must be kept to recover all ground states;
either sign suffices to recover one.

\begin{proposition}[Exact elimination and recovery]
\label{prop:one-spin}
Let $H^{\setminus a}$ be defined by Eq.~\eqref{eq:reduced-H}. Then:
(i) $H^{\setminus a}$ and $H$ have the same minimum energy;
(ii) every minimizer of $H^{\setminus a}$ lifts through
Eq.~\eqref{eq:backsub} to a minimizer of $H$;
(iii) the restriction of every minimizer of $H$ to
$\mathbf{s}_{\setminus a}$ minimizes $H^{\setminus a}$.
\end{proposition}

\begin{proof}
Equation~\eqref{eq:reduced-H} holds for each fixed
$\mathbf{s}_{\setminus a}$; minimizing both sides over
$\mathbf{s}_{\setminus a}$ gives (i). Equation~\eqref{eq:backsub} attains
the inner minimum, so a minimizer of $H^{\setminus a}$ extended by
$s_a^\star$ has energy $\min H^{\setminus a}=\min H$, which is (ii). For
(iii), if $\mathbf{s}$ is a ground state of $H$, then
\begin{equation*}
\min H^{\setminus a}
\leq
H^{\setminus a}(\mathbf{s}_{\setminus a})
\leq
H(\mathbf{s})
=
\min H,
\end{equation*}
and the first and last terms are equal by (i), so both inequalities are
equalities.
\end{proof}

\emph{Writing the reduced interaction as spin products.}
The function
\begin{equation}
\label{eq:local-reduced-function}
F_a(\mathbf{s}_{\mathcal{N}_a})
=
-\left|P_a(\mathbf{s}_{\mathcal{N}_a})\right|
\end{equation}
must be written as a sum of spin products for the solver.
Relabel the $d_a=|\mathcal{N}_a|$ spins of $\mathcal{N}_a$ as
$s_1,\ldots,s_{d_a}$. Every
real-valued function of $d_a$ spins has a unique expansion in products
in which each spin appears at most once:
\begin{equation}
\label{eq:walsh-expansion}
F_a(s_1,\ldots,s_{d_a})
=
\sum_{U\subseteq\{1,\ldots,d_a\}}
\widehat{F}_a(U)
\prod_{j\in U}s_j,
\end{equation}
with coefficients
\begin{equation}
\label{WH}
\widehat{F}_a(U)
=
\frac{1}{2^{d_a}}
\sum_{\boldsymbol{\sigma}\in\{-1,+1\}^{d_a}}
F_a(\boldsymbol{\sigma})
\prod_{j\in U}\sigma_j .
\end{equation}
Each new term contains $|U|\leq d_a$ spins: the constant
$\widehat{F}_a(\emptyset)$ shifts the energy, terms with $|U|=1$ give
local fields on the neighbours, and terms with $|U|\geq2$ give
interactions among them.

In the following result, odd and even refer to reversing all
neighbouring spins together.

\begin{corollary}[Parity of the induced terms]
\label{cor:parity}
If $P_a$ is an odd function of the neighbouring spins, which holds exactly
when every nonzero term that contains $s_a$ has even order (for example, a spin
without a local bias in a pairwise Hamiltonian), then $F_a=-|P_a|$ is
even and $\widehat F_a(U)=0$ for every $U$ of odd size. Elimination then
creates no linear terms and no odd-order interactions; with three
neighbours, it creates only a constant and pairwise terms.
\end{corollary}

\begin{proof}
Every nonzero monomial of $P_a$ has degree $|S|-1$, so $P_a$ is odd exactly
when every such $|S|$ is even. Then
$F_a(-\boldsymbol{\sigma})=F_a(\boldsymbol{\sigma})$, and pairing
$\boldsymbol{\sigma}$ with $-\boldsymbol{\sigma}$ in Eq.~\eqref{WH}
cancels every coefficient with odd $|U|$.
\end{proof}

\emph{Example: a spin coupled to three neighbours.}
Consider a spin $s_a$ with unit couplings to three neighbours and no local
field. The terms containing this spin are $s_a(s_b+s_c+s_d)$, and
Walsh elimination gives
\begin{equation}
\label{eq:cubic-unit-gadget}
s_a(s_b+s_c+s_d)
\quad\longrightarrow\quad
-\left|s_b+s_c+s_d\right|
=
-\frac{3}{2}
-
\frac{1}{2}
\left(
s_bs_c+s_bs_d+s_cs_d
\right).
\end{equation}
The three couplings to the removed spin become a triangle with weights $-\tfrac12$,
so the Hamiltonian remains quadratic; the removed spin is recovered
using $s_a=-\sgn(s_b+s_c+s_d)$. This is the star--triangle rule used in
sparse spin-glass reductions
\cite{boettcher2008reduction,boettcher2024physics}. The Supplementary
Information gives the formulas for weighted couplings at low degree
and a worked example of the transform.

\begin{figure}[!htbp]
\centering
\widefiguregraphic{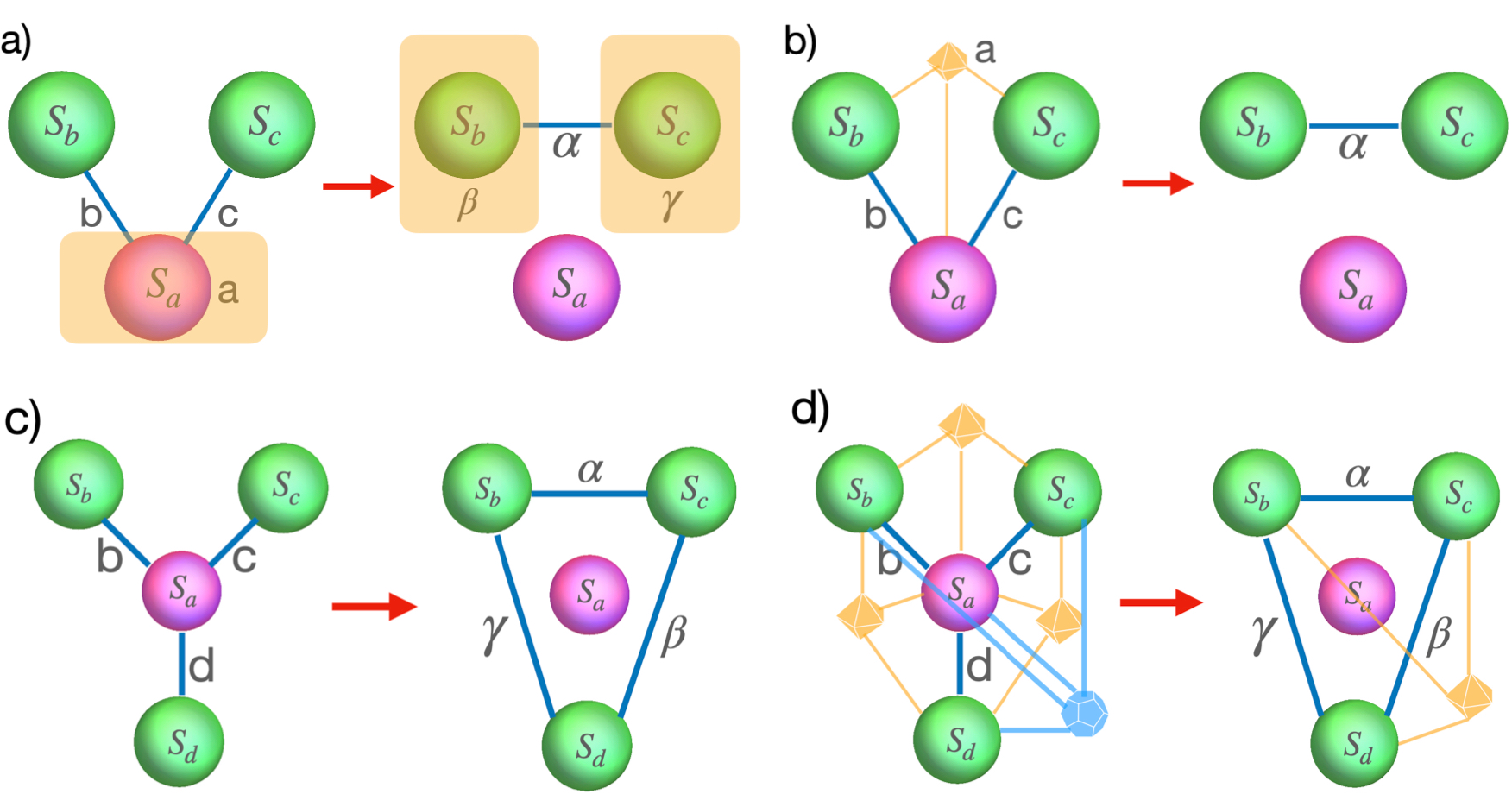}
\caption{%
\textbf{Exact removal of one spin, with a rule to recover it.}
Each red arrow replaces the terms containing $s_a$ by their minimum
over that spin. Magenta marks $s_a$, green its neighbours, and blue
lines pairwise couplings. Amber and light-blue symbols join three
and four interacting spins, respectively; shaded boxes mark local
fields. Latin letters label original coefficients and Greek letters
label coefficients created by elimination.
\textbf{a}, With two neighbours, a local field on $s_a$ can produce
both a pairwise coupling and local fields on the remaining spins.
\textbf{b}, When only two neighbours remain, eliminating pairwise
and three-body terms leaves at most a coupling between those neighbours
and local fields.
\textbf{c}, Three pairwise couplings and no local field give the
weighted star--triangle reduction: pairwise couplings among the
three neighbours, with no local fields.
\textbf{d}, Terms involving up to four spins are replaced by terms
involving at most three.
Possible local fields in \textbf{b,d} and all constant energy offsets
are retained but not drawn; coefficients created by elimination can be zero.
Writing the removed terms as $s_aP_a$, the rule $s_a=-\sgn P_a$
restores an original configuration of equal energy for every reduced
configuration; either sign is allowed when $P_a=0$.
The detached magenta symbol represents this stored recovery rule,
not an active spin.%
}
\label{fig:elimination-schematic}
\end{figure}

Figure~\ref{fig:elimination-schematic} shows four sets of terms containing
the selected spin and the interactions created when it is removed.
Panel~a shows the local fields created when a field on the removed spin
breaks the symmetry of Corollary~\ref{cor:parity}. Panel~c shows the
weighted version of the star--triangle rule in
Eq.~\eqref{eq:cubic-unit-gadget}.
Panels~b and~d illustrate the bound $|U|\leq d_a$: the new terms
involve fewer spins than the largest terms in the original local blocks.
Elimination can instead raise the order when the removed spin has more
neighbours than there are spins in the largest original term.

\emph{Repeated elimination.}
Let $E\subseteq[N]$ be the set of eliminated spins and $R=[N]\setminus E$
the set of remaining spins, and write
$\mathbf{s}=(\mathbf{s}_E,\mathbf{s}_R)$. Applying Eq.~\eqref{eq:reduced-H}
one spin at a time gives
\begin{equation}
\label{eq:partial-min}
H^{\setminus E}(\mathbf{s}_R)
=
\min_{\mathbf{s}_E}H(\mathbf{s}_E,\mathbf{s}_R).
\end{equation}
For a fixed order, $H^{[j]}$ denotes the Hamiltonian after $j$
eliminations, with $H^{[0]}=H$ and $H^{[|E|]}=H^{\setminus E}$.

\begin{corollary}[Order independence for a fixed set of removed spins]
\label{cor:composition}
For each fixed $\mathbf{s}_R$, $H^{\setminus E}(\mathbf{s}_R)$ is the
exact minimum of $H$ over the eliminated spins. For a fixed set $E$, the
final reduced function does not depend on the elimination order.
\end{corollary}

\begin{proof}
Finite minimizations commute, and the multilinear representation of a
function on $\{-1,+1\}^{|R|}$ is unique, so the final reduced function is
independent of the order even though the intermediate polynomials and stored
reverse rules can depend on it.
\end{proof}

Each step stores its local polynomial $P_a$ and the rule of
Eq.~\eqref{eq:backsub}. For a fixed elimination order, we define the
recovery map $\mathcal{L}_E$ by applying these rules in reverse order
and choosing $s_a=+1$ whenever the stored $P_a$ vanishes. The map
returns the full configuration, including the retained spins; the
following result also holds for any other fixed tie rule.

\begin{corollary}[Recovery preserves energy for every configuration]
\label{cor:pointwise-lift}
For every assignment $\mathbf{s}_R$ of the remaining spins,
\begin{equation}
\label{eq:pointwise-lift}
H\!\left(\mathcal{L}_E(\mathbf{s}_R)\right)
=
H^{\setminus E}(\mathbf{s}_R)
=
\min_{\mathbf{s}_E}H(\mathbf{s}_E,\mathbf{s}_R).
\end{equation}
Consequently, for every threshold $\varepsilon\in\mathbb{R}$,
\begin{equation}
\label{eq:sublevel-projection}
\left\{
\mathbf{s}_R:
H^{\setminus E}(\mathbf{s}_R)\leq\varepsilon
\right\}
=
\pi_R\!\left(
\left\{
\mathbf{s}:
H(\mathbf{s})\leq\varepsilon
\right\}
\right),
\end{equation}
where $\pi_R$ restricts a full configuration to the spins in $R$.
\end{corollary}

\begin{proof}
At each reverse step, the stored rule selects a value of the restored spin
that attains the conditional minimum used in the corresponding forward
step. Applying this equality inductively in reverse elimination order gives
Eq.~\eqref{eq:pointwise-lift}. A reduced configuration in the left-hand set
therefore has a full lift with the same energy, so it belongs to the
projection on the right. Conversely, if a full configuration with restriction
$\mathbf{s}_R$ has energy at most $\varepsilon$, then its conditional minimum
at $\mathbf{s}_R$ is also at most $\varepsilon$. Its restriction therefore
belongs to the left-hand set, which proves the reverse inclusion.
\end{proof}

Thus, in exact arithmetic, recovering the removed spins adds no energy
error even when the reduced solver returns a suboptimal configuration.
Equation~\eqref{eq:sublevel-projection} states the same guarantee for any
energy threshold: a reduced configuration meets the threshold exactly
when some original configuration with those retained spins meets it.
Methods, Section~\ref{sec:methods-compiler}, describes rounding, removal
of small coefficients, and checks that the computed expansion reproduces
the local energy table.

\emph{Computational cost.}
To compute the coefficients, we first tabulate $F_a$ on all $2^{d_a}$
neighbour configurations. Let
\begin{equation}
\label{eq:incident-term-count}
r_a
=
\left|
\left\{
S:
a\in S,\;
c_S\neq0
\right\}
\right|
\end{equation}
be the number of terms containing $s_a$, called the incident-term count.
Each table entry sums $r_a$ spin products of at most $k-1$ factors, where
$k$ is the maximum interaction order of the current Hamiltonian, so the
table costs $\mathcal{O}(kr_a2^{d_a})$ operations. Reusing previously
computed spin products can reduce the constant factor. A fast Walsh--Hadamard transform (FWHT)
can compute all $2^{d_a}$ coefficients in $\mathcal{O}(d_a\,2^{d_a})$
operations~\cite{fino1976unified}, giving a total of
$\mathcal{O}((kr_a+d_a)2^{d_a})$ operations to build the table and
compute its Walsh coefficients.

Both the work needed for the transform and the number of coefficients
it can produce grow exponentially with $d_a$. A single step can
produce $2^{d_a}$ terms of order at most $d_a$, adding connections
and widening the range of coefficient values. These costs determine
which spins can be removed within the available resources.

For a given sequence of accepted eliminations, an FWHT and updates
restricted to the affected terms give a cost proportional to the number
of eliminations. This bound assumes fixed limits on the neighbours and
terms of each selected spin and on interaction order (see Supplementary
Information). It excludes choosing the sequence, trying transforms to
check the output limits, and solving the reduced problem.
The benchmark code instead sums the Walsh formula directly, requiring
$\mathcal{O}((d_a+1)4^{d_a})$ operations to compute the coefficients.
It also scans and copies the current Hamiltonian. The measured
preprocessing time includes all these operations; the bound for the
FWHT does not describe the runtime of this code.

With resource limits in place, changing the elimination order can
change which spins remain eligible and hence the set $E$ removed.
Corollary~\ref{cor:composition} applies to each fixed $E$: it does
not say that every order removes the same spins. The number and order
of terms, the range of coefficient values, and the time needed at each
step can still depend on the sequence.

\subsection{Choosing which spins to remove}
\label{sec:compiler}

Removing spins helps only if the remaining problem is practical to solve.
The schedule therefore controls both the cost of elimination and the
interactions it creates. At each step, resource limits determine which
spins are eligible, and an ordering rule chooses which eligible spin to
remove next.

Two local counts guide these choices. The number of neighbours $d_a$
determines the size of the Walsh table and bounds the interaction order
that elimination can create. The number of terms containing the spin,
$r_a$, determines the work needed to evaluate each table entry. In a pairwise model without local fields, both counts equal the number
of neighbours. In a higher-order model, however, even a few terms can
involve many neighbours. Since each term contains at most $k-1$ other spins,
\begin{equation}
\label{eq:da-ra-bound}
d_a\leq(k-1)r_a,
\end{equation}
where $k$ is the current maximum interaction order.

The compiler takes a target fraction $f$ of spins to remove or a target
spin count, together with limits on the interactions created. The
fraction $f$ is measured against the original number of spins, not the
number still present at a later step. Before attempting an elimination,
the code can cheaply check two limits: at most $d_{\mathrm{cap}}$
neighbours and at most $r_{\mathrm{cap}}$ terms containing the selected
spin. Other limits can restrict the order and total number of terms,
or the range of coefficient values. Checking the actual terms created
requires a trial transform, because some Walsh coefficients can be zero.
Limiting the number of neighbours gives a cheaper, conservative bound
on the order of the new terms.

The code maintains a neighbour graph, called the fill graph, that
retains edges even when their coefficients cancel to zero. Its degree
therefore bounds the actual number of neighbours. For quadratic inputs,
\texttt{min\_degree} selects the eligible spin with the lowest degree
in this stored graph. For higher-order inputs, the same named rule
instead selects the eligible spin occurring in the fewest terms, subject
to $r_a\leq r_{\mathrm{cap}}$ and at most $16$ neighbours in the stored
graph. The Supplementary Information describes the alternative
\texttt{min\_fill} and random orders, how the stored graph is updated,
and how spins are checked for eligibility. None of these
rules guarantees the smallest possible elimination width.

The process stops when the target is met or the ordering rule has no
further eligible spin to select. It returns the current reduced Hamiltonian
and the recovery rules. It also reports the fraction of spins removed,
the largest neighbourhood, the order and number of terms, the range
of coefficient values, and the preprocessing time. The input is returned unchanged only if no
elimination was accepted. The algorithms in the Supplementary
Information give the elimination step and scheduling loop.

The benchmarks use \texttt{min\_degree} with
$d_{\mathrm{cap}}=\Delta+1$ for quadratic inputs and
$r_{\mathrm{cap}}=\Delta+1$ for higher-order inputs as low-cost defaults,
where $\Delta$ is the degree of the input regular graph or hypergraph.
The measurements show the limits of this choice. A separate test tracks
how interaction order changes during elimination with minimum-degree
ordering. With $d_{\mathrm{cap}}=12$ and no limits on the number of
terms containing a spin or the order of new terms, random 3-regular
graphs remain pairwise until about $0.37$ of the spins are removed
(Supplementary Fig.~1). At $\Delta=5$, the default
cap stops elimination earlier, after removal of about $0.30$ of the spins
(Supplementary Fig.~2a). Near $\Delta\simeq7$--$8$, the new
terms involve more spins and become much more costly to update,
and the benefit disappears in the
tested settings (Section~\ref{sec:sa-discussion}; Supplementary
Note~3).

\subsection{Quadratic benchmark: sparse Gaussian spin glasses}
\label{sec:sa}

We test whether spin elimination improves solution quality within the
same total time budget by applying simulated annealing to the original
and reduced Hamiltonians with OpenJij~\cite{openjij}. The instances use
Gaussian couplings and no local fields. Each spin has exactly $\Delta$
neighbours in a random regular graph, with $\Delta\in\{3,5\}$ and
$N\in\{100,200,500,1000\}$.
Each formulation receives a total time budget
$T\in\{0.5,1,2,5\}$~s per attempt, measured as elapsed time.
An attempt is a full run within this budget, not a single returned
sample; each attempt can return multiple samples. For every reduced
attempt, we use the elimination settings selected in the development
parameter sweep: $f=0.4$, aiming to remove $40\%$ of the original spins,
\texttt{min\_degree} ordering, and the neighbour cap
$d_{\mathrm{cap}}=\Delta+1$ (Section~\ref{sec:compiler}). The cap can stop elimination before
this target is reached. The full preprocessing time is deducted from each reduced attempt's
budget, leaving the remaining time for annealing.

For $\Delta=3$ and $N\leq200$, we obtain reference minima by
tensor-network contraction. In this calculation, local energy tables
are combined and minimized over shared spin variables until the
minimum energy of the full system is obtained. The algorithm is
exact in exact arithmetic, but our calculations use double-precision
floating-point arithmetic (binary64), which is subject to rounding.
We therefore use ``certified'' to mean numerically verified by an
exact algorithm, with independent checks of the energies of the
returned states, rather than proved with an exact-arithmetic
certificate (Supplementary Note~3.8).

For larger instances, the reference is the lowest energy found by
additional search, including the benchmark runs themselves, and is
not a certified minimum. These comparisons show how closely each formulation approaches the
best energy found across the searches. They do not test performance
against targets fixed independently of the benchmark results
(Supplementary Note~3.8).

\subsubsection{Success rates and energy gaps}
\label{sec:sa-scaling}

An attempt succeeds when it reaches the reference energy within the
$10^{-4}$ tolerance defined in Methods. For the certified setting
$\Delta=3$, $N=200$ and $T=1$~s, elimination raises the fraction
of successful attempts, averaged over instances, from $0.095$
to $0.950$. We denote this average by $\bar p_{\mathrm{succ}}$
and call it the pooled success fraction. Applying
Eq.~\eqref{eq:tts90} to this average gives the time-to-solution
index $\TTSindex$. Elimination reduces this index by a factor
of $30$, with a $95\%$ confidence interval of $[15.7,105.0]$
from paired bootstrap resampling. This index summarizes the tested set, but does
not give the time required to solve any particular instance.

\begin{figure}[!htbp]
\centering
\widefiguregraphic{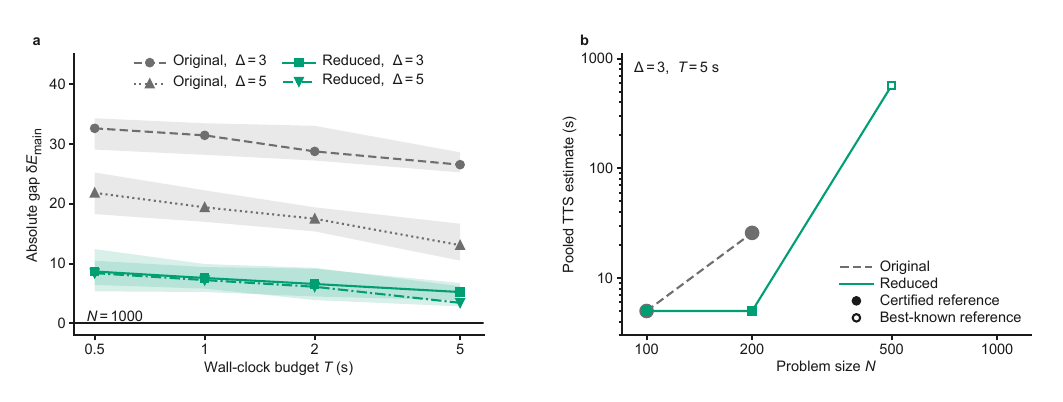}
\caption{%
\textbf{Improved search at fixed time budgets on quadratic spin glasses.}
Every reduced attempt includes the full preprocessing cost.
\textbf{a}, Energy gap
$\delta E_{\mathrm{main}}=E_{\mathrm{best}}-\Erefmain$ at $N=1000$
as the total time budget $T$ increases. Lines show medians over five
instances at each degree $\Delta$; bands span the middle half of
the results (the interquartile range).
The references are best-known energies, not certified minima.
Neither formulation reaches them, but elimination lowers the gaps
at both $\Delta=3$ and $\Delta=5$.
\textbf{b}, Time-to-solution index $\TTSindex$ at $\Delta=3$
and $T=5$~s (Eq.~\eqref{eq:tts90}), plotted on a logarithmic axis.
The index uses the fraction of successful attempts averaged across instances;
it is not a time to solution for each instance.
Grey circles denote the original formulation and green squares the
reduced formulation. Filled markers use certified minima
($N=100,200$); open markers use best-known references ($N=500,1000$).
No finite index is shown when no attempt reaches the reference:
the original formulation at $N=500,1000$ and the reduced formulation
at $N=1000$. With success in every recorded attempt, the plot
shows one full budget ($5$~s), not a confidence bound.
The two markers coincide at $N=100$
(Supplementary Note~3.8).%
}
\label{fig:sa_scaling}
\end{figure}

For larger instances, reaching the reference energy becomes rare.
At $\Delta=3$ and $N=500$, the original formulation records no
success at any tested budget, whereas the reduced formulation
reaches the best-known reference in 2 of 100 attempts at $T=5$~s.
At $N=1000$, neither formulation reaches its reference within the
tested budgets. Nevertheless, at $T=5$~s elimination lowers the
median gap between the best energy returned and the reference energy
from $26.5$ to $5.2$ for $\Delta=3$
and from $13.1$ to $3.4$ for $\Delta=5$
(Fig.~\ref{fig:sa_scaling}a). These gaps are measured from
best-known energies to which the benchmark runs themselves
contributed, not from certified minima.

Generating more samples does not explain all the gains.
At $\Delta=3$, the reduced formulation returns about $1.5$ times
as many samples per second as the original. At $\Delta=5$ and
$N\geq500$, however, it returns only $55$--$66\%$ as many,
yet still reaches the best-known reference more often at
$N=500$, $T=1$~s. Supplementary
Note~3 gives the results for each setting.

\subsubsection{Comparison with variable fixing}
\label{sec:sa-comparison}

We compare Walsh elimination with SPVAR on $60$ instances generated
with new random seeds. For SPVAR, we use the published procedure for
problems without local fields, with multiple starts
\cite{karimi2017effective}. SPVAR uses sampled configurations to choose
spin values before searching the remaining problem. Walsh elimination
instead stores rules that recover the removed spins from the retained
configuration, without fixing their signs before the search.
We also apply roof duality, which removes no spins in these tests;
its annealing runs therefore show how results can differ even when
the Hamiltonian is unchanged.

We compare the methods in three ways: with equal total sample counts,
with equal total time budgets of $20$~s, and with equal total time
budgets of $60$~s.
The tests at the two time budgets reuse the same sequences of random
seeds, so they are not independent repetitions. The primary comparison
uses the best energies found on each instance, without requiring
a reference minimum:
\begin{equation}
\label{eq:paired-energy-difference}
\Delta E_{\mathrm{pair}}
=
E_{\mathrm{method}}-E_{\mathrm{baseline}},
\end{equation}
where $E_{\mathrm{method}}$ and $E_{\mathrm{baseline}}$ are the
best energies returned with and without preprocessing, respectively.
Both are evaluated in the original Hamiltonian, after recovery of
any removed spins. Differences below $-10^{-4}$ count as wins
for preprocessing, those above $10^{-4}$ as losses, and the rest
as ties. Best-known reference energies fixed before these runs
are used only for the secondary comparison of how often each
method reaches a target
(Supplementary Note~3.9).

In these models, every ground state has a partner with all signs
reversed, because $H(s)=H(-s)$. No spin can therefore have the same
sign in every ground state, as strong-persistency fixing
requires~\cite{boros2002pseudo}.
Weak persistency requires only that the fixed values preserve at least
one ground state. Symmetry does not rule out this weaker fixing, but
the tested procedure fixes no spins. The subsequent annealing runs
use the same Hamiltonian as the baseline but different random seeds. Their wins and losses
thus show variation between searches on the unchanged problem,
not an effect of variable fixing.

In $98.8$--$99.5\%$ of SPVAR starts, all fixed spin values agree
with the stored reference configuration or with all its signs reversed.
This observed agreement does not guarantee preservation of a
ground state: an incorrect combination of fixed values can exclude
them all. Under the equal-sample protocol, each start uses $350$
of its $500$ samples in the two stages that select and fix spins,
leaving $150$ for the final search on the remaining problem.
The best energy is retained across all three stages, so the
samples used to make fixing decisions also contribute to the
reported result. SPVAR's win/tie/loss counts against the baseline
are $17/33/10$ at equal samples, $25/30/5$ at $20$~s, and
$17/33/10$ at $60$~s. The sign test detects an improvement only
at the tighter wall-clock budget of $20$~s.

\begin{figure}[!htbp]
\centering
\widefiguregraphic{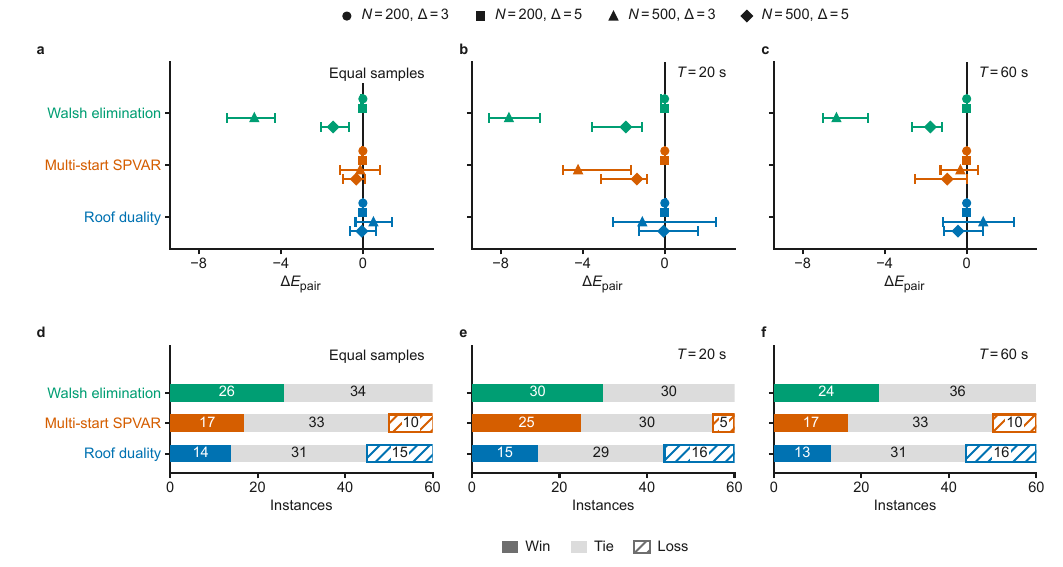}
\caption{%
\textbf{Walsh elimination improves energies without an observed loss.}
The comparison uses $60$ new instances: $20$ at each $N=200$ setting
and $10$ at each $N=500$ setting, with $\Delta\in\{3,5\}$.
Columns compare equal sample counts, equal $20$~s budgets, and equal
$60$~s budgets, respectively.
\textbf{a--c}, For each instance, we subtract the best energy returned
without preprocessing from the best energy returned with preprocessing:
$\Delta E_{\mathrm{pair}}=E_{\mathrm{method}}-E_{\mathrm{baseline}}$.
Each marker shows the median of these differences for one
$(N,\Delta)$ setting, with a $95\%$ confidence interval from paired
bootstrap resampling. Negative differences favour preprocessing. No reference minimum is needed. Marker shapes
identify settings; a marker alone denotes a zero-width interval.
\textbf{d--f}, Numbers of wins, ties and losses against the unreduced
baseline; differences within $\pm10^{-4}$ count as ties.
Walsh elimination records no losses and improves on the baseline
in all three protocols (two-sided sign-test $p_{\mathrm{sign}}<10^{-6}$
in each). SPVAR (sample-persistence variable reduction), run with
multiple starts, shows a statistically detectable improvement only
at $20$~s.
Roof duality fixes no spins: its wins and losses compare runs with
different random seeds on the unchanged Hamiltonian.
The $20$~s and $60$~s tests reuse the same seed streams and are
not independent repetitions. Supplementary
Fig.~3 gives results for each setting;
Supplementary Table~6 shows how often each
method reaches its reference energy.%
}
\label{fig:confirmation}
\end{figure}

Walsh elimination removes $30$--$40\%$ of the spins
(Supplementary Fig.~2a) and records no loss against the baseline
under any of the three protocols. Its win/tie/loss counts are
$26/34/0$ at equal samples, $30/30/0$ at $20$~s, and $24/36/0$
at $60$~s, with two-sided sign-test $p_{\mathrm{sign}}<10^{-6}$
in each case (Fig.~\ref{fig:confirmation}; Supplementary Fig.~3).
At $N=500$, $\Delta=3$, it returns lower energies on all ten
instances under the equal-sample protocol. The median difference is
$\Delta E_{\mathrm{pair}}=-5.29$, with a $95\%$ confidence interval
of $[-6.65,-4.27]$ from paired bootstrap resampling.

A separate validation on $100$ instances gives the same qualitative
pattern. Walsh elimination again records no losses, while SPVAR's
excess of wins over losses decreases as the time budget grows.
At $60$~s, this excess is small, much like the excess of wins or losses
between runs on the unchanged Hamiltonian after roof duality
(Supplementary Table~7).

Walsh elimination therefore improves search under all three tested
protocols, with no observed losses, whereas SPVAR shows a statistically
detectable improvement only at $20$~s and roof duality removes no spins.
The reduction does not commit to spin values inferred from samples:
in exact arithmetic, it preserves the minimum energy, and the stored
rules recover an original configuration of equal energy from every
reduced configuration.

\subsubsection{Robustness checks and limits}
\label{sec:pt}

\paragraph{Checking the reference energies.}
\label{sec:reference-hardening}
The certification calculations reproduce every stored reference energy
for $\Delta=3$ and $N\in\{100,200\}$. For larger instances, additional
search finds lower energies for $30$ of the $70$ references, with
$24$ of these improvements first found through the reduced formulation.
These lower energies give more demanding targets for the comparison,
but remain best-known values rather than certified minima.

\paragraph{Parallel tempering.}
The advantage also appears with parallel tempering. These tests use
$16$ replicas, or copies of the system at different temperatures,
with the same temperature set for both formulations. At
$(N,\Delta)=(200,3)$, elimination lowers the time-to-solution index
by factors of $8.7$--$18.4$. At $(500,3)$, only the reduced
formulation reaches the best-known reference within the tested
budgets. At $(500,5)$ and $T=5$~s, the index falls by a factor
of $9.4$. The gain at $\Delta=5$ persists despite fewer spin
updates per second, so it cannot be explained by a higher update rate.

\paragraph{Preprocessing cost.}
Median preprocessing time is $0.010$--$0.020$~s at $N=200$
and $0.14$--$0.27$~s at $N=1000$. We deduct the full preprocessing
time $t_{\mathrm{prep}}$ from every reduced attempt's budget $T$,
leaving $T-t_{\mathrm{prep}}$ for the solver. Preprocessing therefore
accounts for a fraction $t_{\mathrm{prep}}/T$ of each attempt's
budget; its cost is not shared across restarts.

At $T=0.5$~s, preprocessing takes $2$--$17\%$ of the budget
for $N=200$--$500$ and $28$--$54\%$ at $N=1000$.
At $T=5$~s, these shares fall to $0.2$--$1.7\%$ and
$2.8$--$5.4\%$, respectively
(Supplementary Table~4).
At $N=1000$, neither formulation reaches the best-known reference
within the tested budgets, so no finite time-to-solution index
can be inferred from these runs. Nevertheless, elimination still
reduces the energy gap after its full preprocessing cost is included
(Fig.~\ref{fig:sa_scaling}a).

\paragraph{Limits at higher graph degree.}
\label{sec:sa-discussion}
Removing spins becomes less useful as the interactions created become
more costly. In tests that increase the graph degree, the advantage
disappears near $\Delta=7$--$8$: elimination creates terms involving
eight or more spins, and the solver returns about ten times fewer
samples per second. In this regime, the cost of the additional interactions
offsets the benefit of searching over fewer spins.

\subsection{Solving higher-order problems directly}
\label{sec:klocal}

We next test elimination on problems that already contain higher-order
interactions. Both the original and reduced Hamiltonians are passed
directly to OpenJij \texttt{sample\_hubo}, without converting them
to quadratic models or adding auxiliary spins. Each formulation
receives the same total time budget, with the full preprocessing
time deducted from the reduced formulation's solver time.

We use three problem families. Family~A consists of random spin
glasses in which each interaction contains $k$ spins and each spin
appears in $\Delta$ interactions. These are mainly three-body
problems ($k=3$), with $\Delta\in\{3,5\}$ and sizes up to $N=513$.
Family~B contains planted 3-XORSAT instances up to $N=128$.
Each parity constraint involves three variables, and each variable
appears in three constraints (3-regular). Family~C contains
low-autocorrelation binary sequence (LABS) problems, whose
interactions are dense. Of the $266$ instances, $136$ have certified
minimum energies and $130$ use best-known references. Each group
of instances with the same family and parameters uses only one
reference type.

The default schedule targets removal of $40\%$ of the original
spins ($f=0.4$). It uses \texttt{min\_degree} ordering, which for
higher-order inputs selects the eligible spin appearing in the
fewest terms, and the cap $r_{\mathrm{cap}}=\Delta+1$, which limits
the number of current terms containing a spin selected for
elimination. The resource limits can stop elimination before
the target fraction is reached
(Supplementary Note~4.4).

\subsubsection{Reaching certified minima}
\label{sec:klocal-success}

For $k=3$, $\Delta=3$, $N=64$ and $T=2$~s, elimination raises
the observed success fraction, averaged over instances, from
$0.122$ to $0.806$. Applying Eq.~\eqref{eq:tts90} to these
averages shows that elimination reduces the time-to-solution index
by a factor of $12.6$, with a $95\%$ confidence interval of
$[8.8,20.3]$ (Table~\ref{tab:klocal-success}).
At the same budget, the index falls by factors of $4.4$ and
$3.5$ on 3-XORSAT at $N=96$ and $N=128$, respectively.
At $N=128$, however, success remains rare: the observed fractions
are $0.008$ for the original formulation and $0.028$ for the
reduced formulation, and the ratio has a wide confidence interval.

Not every setting shows a clear gain. For $k=3$, $\Delta=5$
and $N=63$, all twenty reference energies were certified after
the benchmark runs. Both formulations reach the certified
minimum in $93$--$95\%$ of attempts at $T=2$~s.
The index ratio is $1.1$, with a $95\%$ interval of
$[0.97,1.26]$, which includes the no-change ratio of one.

The main gain also appears with a second solver that accepts
higher-order interactions directly. With \texttt{qubovert}, elimination
reduces the index by a factor of $9.9$ for the $k=3$, $\Delta=3$,
$N=64$ setting; the $95\%$ interval is $[6.4,17.8]$.
An OpenJij rerun on the same machine gives a factor of $12.1$,
with an interval of $[7.6,22.4]$
(Supplementary Table~9).

\begin{table}[t]
\centering
\caption{%
\textbf{Success at certified minima with preprocessing included.}
Each attempt has a total budget of $T=2$~s.
$\bar p_{\mathrm{succ}}$ is the fraction of successful attempts averaged
across instances. The index ratio divides $\TTSindex$ for the original
formulation by its value for the reduced formulation
(Eq.~\eqref{eq:tts90}). It uses these averaged success fractions,
not the time to solve each instance.
The $95\%$ intervals use paired bootstrap resampling over instances.
Bold ratios have a lower interval bound above the pre-specified
$3\times$ reporting threshold.
$^{\dagger}$~Minima certified after the benchmark runs, included
under the reporting rule in Supplementary
Note~4.4.%
}
\label{tab:klocal-success}
\small
\begin{tabular}{@{}l r r r l@{}}
\toprule
Problem & \multicolumn{2}{c}{$\bar p_{\mathrm{succ}}$}
    & Index ratio & $95\%$ interval \\
\cmidrule(lr){2-3}
    & Original & Reduced & & \\
\midrule
$k=3$, $\Delta=3$, $N=64$
    & 0.122 & 0.806 & $\mathbf{12.6\times}$ & $[8.8,20.3]$ \\
3-XORSAT, $N=96$
    & 0.122 & 0.436 & $\mathbf{4.4\times}$ & $[3.6,5.6]$ \\
3-XORSAT, $N=128$
    & 0.008 & 0.028 & $3.5\times$ & $[1.8,14.2]$ \\
$k=3$, $\Delta=5$, $N=63$\,$^{\dagger}$
    & 0.930 & 0.948 & $1.1\times$ & $[0.97,1.26]$ \\
\bottomrule
\end{tabular}
\end{table}

\subsubsection{Energy gaps on larger instances}
\label{sec:klocal-gap}

When too few attempts reach the reference energy for a useful
comparison of success fractions, we measure how close the returned energies
are to that reference. We use the relative energy gap
\begin{equation}
\label{eq:klocal-relative-gap}
g_{\mathrm{HO}}
=
\frac{E_{\mathrm{best}}-\Erefho}{|\Erefho|},
\end{equation}
where $E_{\mathrm{best}}$ is the best energy returned and
$\Erefho$ is the reference energy for the same instance.
We report this gap as a percentage. For the large spin-glass
instances, the references are best-known energies, so these gaps
measure distance from the best values found, not from certified
minima (Supplementary Note~4.4).

At $k=3$, $\Delta=3$ and $T=30$~s, elimination lowers the
median relative gap from $1.29\%$ to $0.15\%$ at $N=256$
and from $1.95\%$ to $0.37\%$ at $N=512$.
These are reductions by factors of $8.7$ and $5.2$, respectively.
The $95\%$ confidence intervals for the differences in gaps, computed
by paired bootstrap resampling, exclude zero in both settings
(Table~\ref{tab:klocal-gap}).

The gains are smaller at $\Delta=5$: the median gap decreases
by a factor of $1.6$ at $N=255$ and $1.2$ at $N=513$.
The interval for the difference in gaps includes zero at $N=255$
but excludes it at $N=513$. Additional search found lower reference
energies for some $N=513$ instances. Thus, a detected difference
between formulations does not make these references certified minima. The smaller gains at
higher degree are consistent with the greater cost of evaluating
the interactions created by elimination.

Supplementary Figure~5 shows how the
gaps change with the time budget at $\Delta=3$. At short budgets,
preprocessing leaves no solver time for some instances, so these
instances are omitted from the reduced curves. The set of
instances contributing to those curves can therefore change with
the budget. A separate check uses the same set of instances at
every budget from $T=2$~s onward and still favours the default
elimination schedule
(Supplementary Fig.~6).

\begin{table}[t]
\centering
\caption{%
\textbf{Smaller energy gaps on larger three-body spin glasses.}
Values are median relative gaps at $T=30$~s, with preprocessing
included. The gap $g_{\mathrm{HO}}$ in
Eq.~\eqref{eq:klocal-relative-gap} measures distance from the
best-known reference, not a certified minimum.
The gap ratio divides the original median by the reduced median.
The last column states whether the $95\%$ confidence interval for
the difference in gaps, computed by paired bootstrap resampling,
excludes zero.
Bold ratios exceed $3\times$ as a visual guide only; the
pre-specified reporting threshold applies to time-to-solution
index ratios, not to gaps.%
}
\label{tab:klocal-gap}
\small
\begin{tabular}{@{}l r r r c@{}}
\toprule
Setting & Original gap & Reduced gap & Gap ratio
    & \shortstack{$95\%$ interval\\excludes zero?} \\
\midrule
$\Delta=3$, $N=256$
    & 1.29\% & 0.15\% & $\mathbf{8.7\times}$ & yes \\
$\Delta=3$, $N=512$
    & 1.95\% & 0.37\% & $\mathbf{5.2\times}$ & yes \\
$\Delta=5$, $N=255$
    & 1.42\% & 0.90\% & $1.6\times$ & no \\
$\Delta=5$, $N=513$
    & 2.50\% & 2.05\% & $1.2\times$ & yes \\
\bottomrule
\end{tabular}

\smallskip
\footnotesize
Additional search lowered the reference energy for $5$ of the
$20$ instances at $\Delta=5$, $N=513$. The references at
$\Delta=3$, $N=512$ did not change under the same additional
search. Neither result certifies optimality: the gaps in both
settings remain comparisons with best-known energies.
\end{table}

\subsubsection{Where elimination stops helping}
\label{sec:klocal-boundary}

At $N=512$, limiting the number of terms containing the selected spin
keeps the median time spent on preprocessing below $2$~s.
Limiting the number of spins in each new term is more costly:
the compiler must try the transform to see which terms would be
created before accepting an elimination. With these checks, preprocessing takes
$11$--$25$~s.

For the tested four-body problems ($k=4$), preprocessing takes
$108$--$170$~s. The reduced model has a higher observed success fraction
once constructed, but this preparation cost makes the original
formulation perform better within the same total time budget.
For dense LABS instances, no spin can be removed under the
tested limits on the number of spins in each term, so the compiler returns the
original problem unchanged
(Supplementary Fig.~4;
Supplementary Table~8).

\subsubsection{Where elimination is most useful}
\label{sec:klocal-discussion}

The largest gains in these tests occur on sparse three-body
problems. The gains decrease as each spin participates in more
interactions, and preprocessing can outweigh the benefit on
four-body problems. Thus, removing more spins helps only when
the cost of constructing and solving the reduced model remains
low enough.

\subsection{Confirmation on new instances}
\label{sec:heldout}

We test whether the gains persist on new instances without further
tuning. We select two settings from the development results:
quadratic spin glasses with $\Delta=3$, $N=200$ and $T=1$~s,
and three-body spin glasses with $k=3$, $\Delta=3$, $N=64$
and $T=2$~s. We fix and record the full protocol before drawing
the random seed from which $40$ new instances are generated in each
setting. For each instance, both formulations use the same solver
routine and share an annealing schedule calculated from the
original Hamiltonian.

We verify each reference minimum with two tensor-network calculations
that combine the local energy tables in independently chosen orders.
Each calculation minimizes over shared spin variables; this is tropical
contraction. The two calculations must agree on the minimum energy,
and we independently evaluate the returned configurations in the original
Hamiltonian. One quadratic instance exceeds the time limit during the
second calculation. It is excluded from analyses that require
a verified minimum, but retained in the comparison of achieved
energies.

We run $50$ attempts per instance and formulation, each with the
total time budget $T$. The full preprocessing cost is included
in every reduced attempt. For each instance with a verified
minimum, we calculate the fraction of successful attempts for each
formulation and subtract the original fraction from the reduced fraction.
The primary measure is the mean of these differences across instances.
The test counts as confirmation only if the entire $95\%$ confidence
interval for this mean lies above zero; this rule is fixed in advance.
We also compare achieved energies and estimate success after repeated
attempts. The time-to-solution index calculated from the mean success
fraction is an additional summary, not the primary test of improvement
(Supplementary Note~6.1;
Supplementary Table~10).

\begin{figure}[!htbp]
\centering
\widefiguregraphic{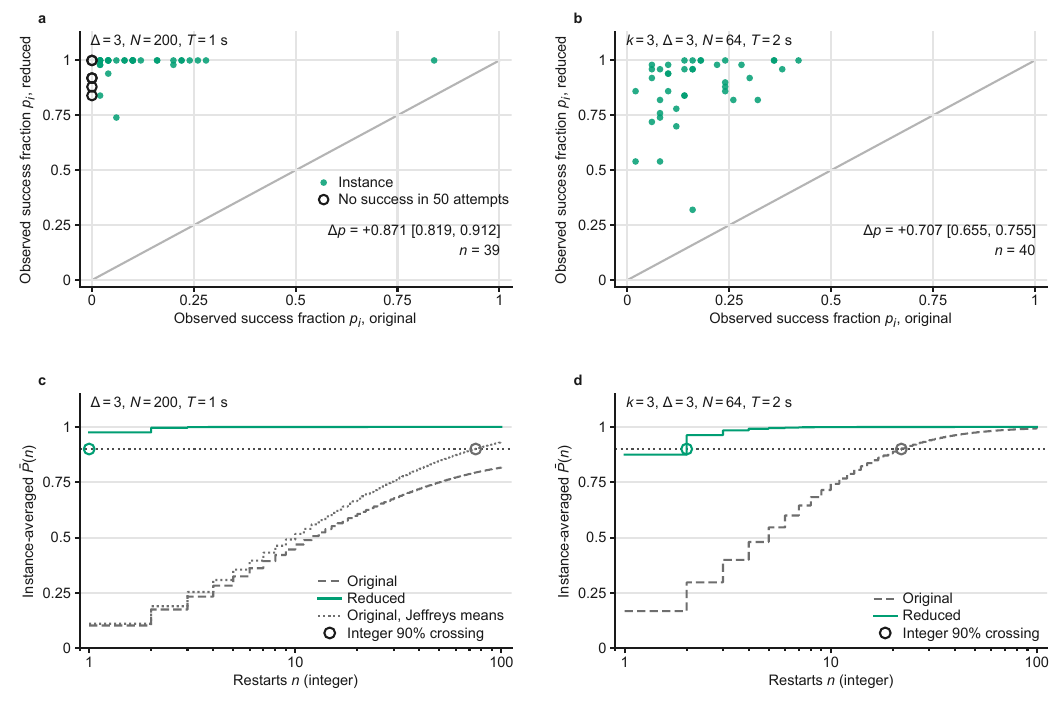}
\caption{%
\textbf{Higher observed success on every verified instance.}
The protocol is fixed before the new instances are generated.
Both formulations share an annealing schedule, and every reduced
attempt includes the full preprocessing cost.
\textbf{a,b}, Each point compares success fractions $p_i=k_i/50$
from $50$ attempts per formulation on the same instance;
$k_i$ counts successful attempts.
All points lie above the equality line; coincident points overlap.
The quadratic test (\textbf{a}) includes $39$ verified instances;
one of the planned $40$ exceeds the time limit during its second
reference verification. The three-body test (\textbf{b}) includes
all $40$ instances. Hollow points mark six quadratic instances where
the original formulation never succeeds in the recorded attempts;
this does not imply that success is impossible. Annotations give
the increase $\Delta p$, averaged over instances, and two-sided
$95\%$ confidence intervals from $20\,000$ bootstrap resamples.
We resample instances, then paired attempts within each selected
instance; $n$ here counts instances.
\textbf{c,d}, Estimated probability of at least one success after
$n$ independent attempts, averaged over instances:
$\bar P(n)=\langle1-(1-p_i)^n\rangle_i$.
Here $n$ counts attempts. Hollow circles on the $0.9$ line mark
the first integer count that reaches this threshold: one for the
reduced quadratic model, and $22$ versus $2$ for the original
and reduced three-body models. The circles mark the threshold,
not the curve values. The original quadratic estimate approaches
$33/39$ and never reaches $0.9$. Replacing $p_i$ by the smoothed
values $(k_i+0.5)/51$ gives the dotted curve, which crosses at
$n=75$. This checks how limited sampling affects the estimate.
It substitutes a single smoothed probability for each instance rather
than propagating uncertainty in that probability through the restart
calculation; it is not a confidence bound.
Grey dashed lines show the original formulation and green solid
lines the reduced formulation. These curves describe average
success, not a guarantee for each instance.%
}
\label{fig:heldout}
\end{figure}

Both settings meet the confirmation criterion
(Fig.~\ref{fig:heldout}). For the quadratic problems, the fraction
of successful attempts, averaged over instances, rises from
$0.104$ to $0.975$. The mean increase is $0.871$, with a $95\%$
interval of $[0.819,0.912]$. For the three-body problems, the
average rises from $0.169$ to $0.875$. The mean increase is
$0.707$, with an interval of $[0.655,0.755]$.
The observed success fraction increases on every one of the
$39$ quadratic and $40$ three-body instances with verified
minima.

Achieved energies also improve. Each attempt returns one energy: the
lowest energy among the samples it produced within its budget,
evaluated in the original Hamiltonian after reverse recovery for the
reduced formulation. For each of the $50$ paired attempts on an
instance, we subtract the original attempt's energy from the reduced
attempt's energy. The instance statistic is the median of these $50$
paired differences, and we report the median of the instance statistics
across instances, with a $95\%$ interval from the same bootstrap over
instances and then paired attempts (Methods,
Eq.~\eqref{eq:heldout-energy}). These medians are
$-0.730$ $[-0.895,-0.536]$ for the quadratic problems and
$-0.126$ $[-0.161,-0.080]$ for the three-body problems. A two-sided
sign test over instances gives $p<10^{-11}$ in both settings.
The energy comparison includes all $40$ quadratic instances,
because it does not require a verified minimum. This statistic differs
from a comparison of the lowest energies found over all $50$ attempts,
whose median difference is $0$ in both settings.

To assess repeated attempts, we first estimate the chance of
at least one success on each instance and then average over
instances (Methods, Eq.~\eqref{eq:pooled-jensen}).
Let $n_{90}$ be the smallest number of attempts for which
this average reaches $0.9$. For the three-body problems,
$n_{90}$ falls from $22$ to $2$ after elimination.
For the quadratic problems, the reduced estimate reaches
$0.9$ after one attempt. The original estimate never reaches
this threshold because six instances have zero observed
successes. The failure of this estimated curve to reach $0.9$ reflects limited
sampling, not evidence that those instances cannot be solved
with more attempts. These curves describe average success
across the tested instances, not a guarantee for each instance.

As a separate summary, we calculate the time-to-solution index from
the mean success fraction. Elimination reduces it by factors of
$33.7$ and $11.3$ for the quadratic and three-body problems,
respectively, close to the development estimates of $30$ and $12.6$.

The reduced quadratic formulation returns more samples per second
($834$ versus $535$). The higher sample rate contributes to performance within the common
time budget, but these measurements do not separate
the benefit of returning more samples from any change in success
per sample. All reported gains include the full preprocessing
cost on every attempt. Sharing this cost across repeated attempts
would change the available solver time by less than $1\%$.
The test therefore confirms the gains on new instances in the
two selected settings, using a shared schedule without further
tuning. The size of the gains can still depend on the machine,
time budget, and annealing schedule.

\FloatBarrier

\subsection{Cubic Max-Cut: larger problems and fewer restarts}
\label{sec:casestudy}
\label{sec:maxcut}
\label{sec:AOC_spin_elimination}
\label{sec:aoc_qumo}
\label{sec:aoc_elim}
\label{sec:aoc_protocol}
\label{sec:aoc_results}
\label{sec:aoc_discussion}

In cubic Max-Cut, each graph vertex has three neighbours; the
Hamiltonian is pairwise, not three-body. For connected simple graphs of this type
with more than four vertices, we prove that elimination removes
at least one-third of the spins without introducing higher-order
interactions or increasing the number of couplings. This allows
inputs up to $50\%$ larger within the same spin budget, provided
that the solver can represent the remaining couplings.

As a concrete example, we use the $16$-variable configuration
demonstrated by the Analog Optical Computer (AOC), giving a spin
budget of $N_{\mathrm{chip}}=16$~\cite{kalinin2025analog}.
The AOC uses optics for matrix--vector multiplication and analog
electronics for the updates. We first test which graphs can be
reduced to this spin count. Separate software tests compare search before and after elimination
using the AOC update rule. These tests include graphs larger than the
$16$-spin budget and do not measure physical hardware performance.

\paragraph{Problem and exact reduction.}
Max-Cut divides the vertices into two groups to maximize the
number of edges between them. Assigning a spin $s_i\in\{-1,+1\}$
to each vertex identifies its group. The cut value $Q$ and
the corresponding Ising energy $H$ are
\begin{equation}
\label{eq:cut_value}
Q(\mathbf{s})
=
\sum_{(i,j)\in E(G)}\frac{1-s_is_j}{2}
=
\frac{|E(G)|-H(\mathbf{s})}{2},
\qquad
H(\mathbf{s})=\sum_{(i,j)\in E(G)}s_is_j.
\end{equation}
Thus, minimizing $H$ maximizes the cut.
By Corollary~\ref{cor:parity}, eliminating a spin with three
neighbours and no local field creates only a constant and
pairwise couplings. Selecting vertices that are not adjacent
to one another lets us repeat this step without changing the
neighbourhoods of the other selected vertices.

\begin{proposition}[Spin and interaction bounds for cubic graphs]
\label{prop:cubic-capacity}
Let $G$ be a connected simple cubic graph with $N>4$ vertices,
and consider a zero-field pairwise Ising Hamiltonian whose
nonzero couplings form $G$. There is a set $I\subseteq V(G)$
of mutually nonadjacent vertices with
$|I|\geq\lceil N/3\rceil$ such that eliminating all spins
in $I$, in any order, creates only constants and pairwise
interactions, with no local fields.

At every step, each remaining spin has at most six neighbours,
and the number of pairwise couplings does not exceed the
original count of $3N/2$. The remaining spin count satisfies
\begin{equation}
\label{eq:cubic-capacity}
N_{\mathrm{red}}
\leq
N-\left\lceil\frac{N}{3}\right\rceil
=
\left\lfloor\frac{2N}{3}\right\rfloor.
\end{equation}
Consequently, every such instance with
$N\leq\lfloor3N_{\mathrm{chip}}/2\rfloor$ can be reduced
exactly to at most $N_{\mathrm{chip}}$ spins while retaining
pairwise interactions and degree at most six.
Once $I$ is given, elimination and storage of the reverse
rules require $\mathcal{O}(N)$ arithmetic operations and
space in a sparse representation.
\end{proposition}

\begin{proof}
Since $G\ne K_4$, Brooks' theorem gives a three-colouring
in which adjacent vertices have different
colours~\cite{brooks1941colouring}.
The largest group of vertices of one colour contains at
least $\lceil N/3\rceil$ vertices, no two of which are
adjacent. Take this group as $I$.

Eliminating a spin in $I$ adds terms only among vertices
outside $I$. It therefore leaves the local block of every
other selected spin unchanged. Each selected spin is
eliminated with three neighbours and no field, so
Corollary~\ref{cor:parity} permits only constants and
pairwise terms. Removing all spins in $I$ gives
Eq.~\eqref{eq:cubic-capacity}.

For any remaining vertex $v$, let $t_v$ count its original
neighbours already eliminated from $I$. Each such elimination
removes one edge at $v$ and can add at most two. Hence
\begin{equation}
\label{eq:cubic-residual-degree}
\deg_{\mathrm{res}}(v)
\leq 3-t_v+2t_v
=3+t_v
\leq6.
\end{equation}
Each elimination removes three edges and adds at most three.
Couplings between the same pair are combined, and cancellations
can only reduce the count. The total therefore never exceeds
$|E(G)|=3N/2$, with the constant stored separately.

Each step uses a table for only $2^3=8$ neighbour configurations,
changes a bounded number of couplings, and stores a fixed-size
reverse rule. Sparse coupling lists therefore give a linear
total cost in operations and storage once $I$ is supplied.
\end{proof}

The argument also allows arbitrary edge weights in pairwise models
without local fields. More generally, suppose the coupling graph can
be coloured with $q$ colours so that adjacent vertices have different
colours, and no vertex has more than $D_G$ neighbours. Eliminating
the largest colour group removes at least $\lceil N/q\rceil$ spins
and creates no local fields. Each new interaction contains an even
number of spins, no more than $D_G$.
For cubic graphs, $q=D_G=3$, so only pairwise interactions
can appear.

These guarantees concern spin count and interaction structure,
not solver time. They do not bound the coupling strengths,
and at most six neighbours per spin does not ensure that
the required connections are available on a particular device.

\paragraph{Testing practical elimination orders.}
We apply four elimination orders to $100$ random cubic graphs
at each even size $N=18,\ldots,40$.
Unlike the proof, these orders select spins one at a time
rather than removing a colour group. A spin is eligible
only when it has at most three neighbours in the graph
stored by the code. This stored graph retains edges whose
weights cancel to zero, so its degree can exceed the number of
nonzero couplings at a vertex. The eligibility check is
therefore conservative and keeps every elimination pairwise
by Corollary~\ref{cor:parity}.

These orders can remove more spins than the colour-group
construction, but impose no degree limit on the remaining
vertices. Separate search tests compare the original and
reduced formulations with the same number of restarts. In every search
test, we restore the removed spins and evaluate the cut in
the original graph
(Supplementary Note~9).

\paragraph{Fitting larger problems into the spin budget.}
Proposition~\ref{prop:cubic-capacity} guarantees that every
graph in the stated class with $N\leq24$ fits a $16$-spin
budget, with no higher-order interactions and degree at
most six. All four tested orders also fit every graph
at each tested size through $N=24$.

\begin{figure}[!htbp]
\centering
\widefiguregraphic{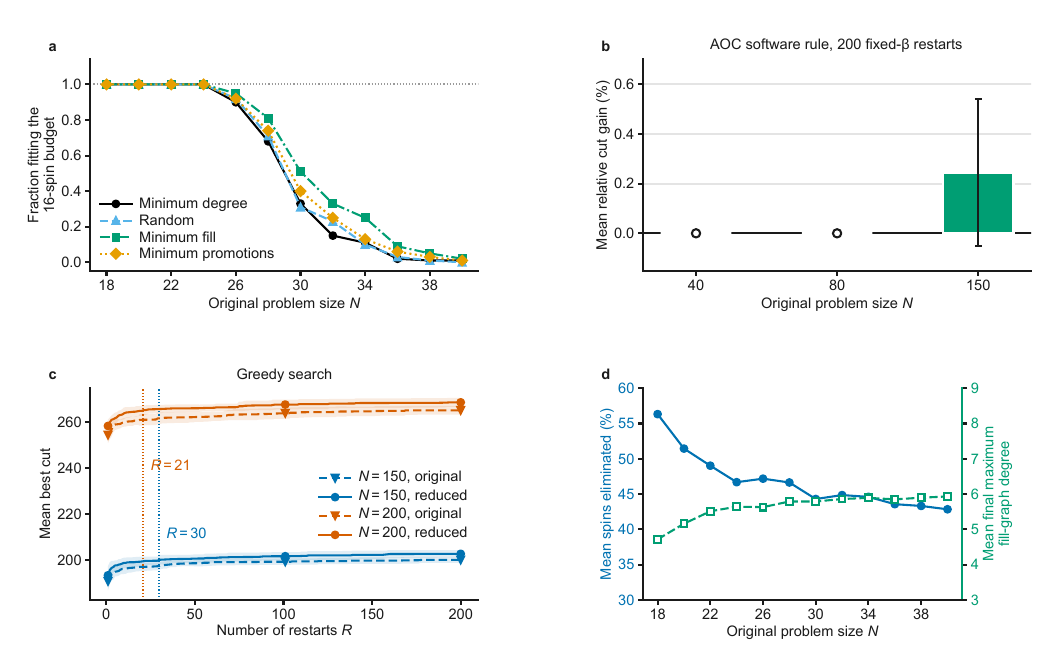}
\caption{%
\textbf{Larger inputs and fewer restarts for cubic Max-Cut.}
All panels show software tests. The $16$-spin budget matches the
Analog Optical Computer (AOC) demonstration; panel~\textbf{b} uses
its fixed-point update rule. These are not measurements on the device
or results from a simulation validated against the complete hardware. Cuts are evaluated in
the original graph after spin recovery.
\textbf{a}, Fraction of $100$ graphs at each even size
$N=18,\ldots,40$ reduced to at most $16$ spins.
All four elimination orders fit every sampled graph through
$N=24$. At $N=32$, minimum fill fits $33\%$ of graphs, compared
with $15\%$ for minimum degree. Minimum degree is an elimination
order, not an unreduced baseline.
\textbf{b}, For each graph, we take the best cut from $200$ restarts
of each formulation under the AOC rule, with $\beta$ fixed.
We calculate the percentage change relative to the original cut
and average it over the $20$ graphs at each size.
Best cuts are unchanged on every graph at $N=40,80$ (open circles).
At $N=150$, the mean gain is $0.245\%$, with $9$ wins, $11$ ties
and no losses.
\textbf{c}, For a separate greedy solver, the best cut from the first
$R$ restarts is found for each graph, then averaged over $20$ graphs.
Restarts follow the saved run order.
The reduced formulation reaches the original formulation's average
best cut from $200$ restarts after $30$ restarts at $N=150$ and
$21$ at $N=200$. The ratios, $6.7$ and $9.5$, compare the numbers
of restarts needed to reach the same average cut. They do not measure
time saved or speedups for individual graphs. Error bars in \textbf{b} and bands in \textbf{c} show one
standard deviation across graphs, not confidence intervals.
\textbf{d}, Percentage of spins removed (left axis, blue circles)
and largest number of neighbours of any remaining spin after
elimination (right axis, green open squares). Both are measured
in each run and then averaged over the $100$ graphs and four
orders at each size. Neighbours are counted in the stored graph,
which retains zero-weight edges and can therefore overcount the
nonzero couplings at a spin. No spin has more than six neighbours
at the end of any run, although these orders do not enforce this limit
(Supplementary Note~9).%
}
\label{fig:AOC_composite}
\end{figure}

Some larger graphs fit as well. At $N=32$,
\texttt{min\_fill} reduces $33\%$ of the graphs to at most
$16$ spins, compared with $15\%$ for deterministic
minimum-degree ordering
(Fig.~\ref{fig:AOC_composite}a,d).
Averaged over the $100$ graphs and four orders at each size,
the removed fraction ranges from $43\%$ to $56\%$ across
$N=18$--$40$.
For each graph and order, we record the largest number of neighbours
of any spin after elimination, using the stored graph. We then average
these maxima over graphs and orders at each size. The averages range
from $4.72$ to $5.93$; no individual maximum exceeds six, although
the tested orders do not enforce this bound.

\paragraph{Search with the AOC update rule.}
We compare the best cuts from $200$ restarts per formulation
on $20$ graphs at each size.
At $N=40$ and $N=80$, both formulations return the same
best cut on every graph. At $N=150$, elimination uses about
$38\%$ fewer spins and returns a better cut on nine graphs,
the same cut on eleven, and a worse cut on none.
Averaging the percentage increase for each graph gives $0.245\%$.
The mean gain in the number of cut edges is $0.50$ per graph.
A paired $t$-test on the differences in cut edges gives
$t_{19}=3.684$ and a two-sided $p=0.00158$, without adjustment
for multiple comparisons.

Figure~\ref{fig:AOC_composite}b uses $200$ restarts per
formulation with the solver parameter $\beta$ held fixed.
A separate test rescales $\beta$ for the reduced model. Including
its results leaves the best cut on every graph unchanged
(Supplementary Note~9).

\paragraph{Reaching the same mean cut with fewer restarts.}
In a separate test with a greedy solver, we take the original
formulation's best cut from $200$ restarts on each graph and
then average over graphs. The reduced formulation reaches this
average after only $30$ restarts at $N=150$ and $21$ at $N=200$.
The original-to-reduced restart-count ratios are therefore $6.7$
and $9.5$ at the same average cut. With all $200$ restarts, elimination
increases the average best cut by $1.32\%$ and $1.30\%$,
respectively. These ratios describe averages across graphs, not
speedups for each graph or reductions in elapsed time.

\FloatBarrier

\section{Discussion}
\label{sec:scope}
\label{sec:conclusions}

A smaller Ising formulation can be substantially easier to solve,
even when its interactions are more complex. Walsh elimination
uses this trade in the opposite direction to quadratization:
instead of adding spins to lower interaction order, it removes
spins while allowing the remaining interactions to change.
In exact arithmetic, this reduction preserves not only the
minimum energy but also the quality of every returned answer:
each reduced configuration can be restored to an original
configuration of equal energy. The solver can therefore search
over fewer variables without committing to guessed spin values
or introducing an approximation into the objective.

The elimination identity and its expression as spin products are
established in pseudo-Boolean optimization and spin-glass reduction
\cite{Crama1990,boros2002pseudo,boettcher2008reduction}.
Min-sum elimination applied to tables produces the same reduced
function for a fixed set of removed spins. Our contribution is a
compiler that applies this exact operation under explicit resource
limits, supplies the polynomial required by higher-order solvers,
and keeps the rules needed to recover the removed spins.
The benchmarks identify settings where this improves search;
the cubic-graph result guarantees a larger input within a fixed
spin budget.

The gains persist on new instances without further tuning. The
confirmation test uses $40$ new quadratic and $40$ new three-body
instances, with its protocol fixed before generation and the same
annealing schedule for both formulations. Elimination increases the
fraction of successful attempts, averaged over instances, by $87$
and $71$ percentage points, respectively. Observed success increases
on all $39$ quadratic and $40$ three-body instances with verified
minima. The remaining quadratic instance exceeds the time limit during
reference verification but remains in the energy comparison
(Section~\ref{sec:heldout}). Every reduced attempt includes the full
preprocessing cost. In the development
tests, elimination reduces the time-to-solution index by factors of
$30$ and $12.6$, with comparable ratios in the confirmation test.
This index is calculated from success fractions averaged across
instances. The advantage also appears with parallel tempering and
a second solver that accepts higher-order terms. In the separate
variable-fixing comparison, Walsh elimination removes
$30$--$40\%$ of the spins and significantly improves on
the unreduced baseline under all three tested protocols,
with no observed losses. Together, these results show
that exact reduction can improve practical optimization,
not merely preserve its answer.

The capacity increase is guaranteed for a whole graph class,
not only for the sampled instances. For pairwise models with no local
fields, Proposition~\ref{prop:cubic-capacity} guarantees removal
of at least one-third of the spins whenever the graph is connected,
simple, and cubic, with more than four vertices. The reduced model
remains pairwise, has no local fields, has at most six neighbours
per spin, and contains no more couplings than the original.
This permits inputs $50\%$ larger within the same spin
budget, provided that the solver supports the required
couplings and their strengths. In particular, every eligible
$24$-spin problem has an exact representation using at most
$16$ active spins. Thus, a larger input need not require
either more solver spins or higher-order interactions.

The benefit ends when preparing the reduced model and evaluating its
new interactions cost more than is saved by removing spins.
Sparse problems give the largest gains; increasing the number of
connections makes the interactions created by elimination more costly, and
preprocessing can outweigh the search benefit on four-body
problems. Exactness alone does not guarantee faster search:
elimination can change local minima, energy barriers,
and the probability distribution of configurations at finite temperature.
A comparison at the same total time budget with classical rules that
remove spins with up to three neighbours would separate the additional
benefit of allowing higher-order terms from that of pairwise reduction
alone.
Further tests should cover more sizes, degrees, and budgets,
reduce the cost of checking the terms created by elimination, and
measure physical implementations with preprocessing, transfer,
control, and readout included.

The design principle is to choose the problem representation for
the solver's resources,
rather than minimize interaction order alone. Higher-order
interactions can be useful when they allow enough spins
to be removed, while suitable graph structure can provide
the same capacity gain within a pairwise model.
Larger problems and more reliable search can therefore
come from an exact change in representation, not only
from building a larger solver.

\section{Methods}
\label{sec:methods}

\subsection{Compiler and numerical checks}
\label{sec:methods-compiler}

The compiler stores each term as a coefficient and the set of spins
in its product. It follows the elimination schedules in
Section~\ref{sec:compiler} and recovers removed spins in reverse
order. Calculations use double-precision floating-point arithmetic
(binary64). Coefficients of the local polynomial $P_a$ are removed
when their magnitude is at most $10^{-15}$. Walsh coefficients
and coefficients obtained by combining terms are kept only when
their magnitude exceeds $10^{-12}$. During recovery, the code
treats $|P_a|\leq10^{-12}$ as a tie and sets $s_a=+1$.
Every recovered configuration is evaluated again in the original
Hamiltonian.

We check each Walsh expansion by reconstructing its local energy
table. The largest discrepancy accumulated during these table checks is
$1.7\times10^{-14}$ in the original certified settings and
$7.1\times10^{-14}$ in the confirmation test, well below the
$10^{-4}$ tolerance used to count successful attempts.
The reconstruction check excludes the effects of removing small
coefficients from $P_a$, discarding small coefficients after
terms are combined, and treating near-zero values as ties
during recovery. These discrepancies measure how well the retained expansion reproduces
the table. They do not bound all numerical errors or establish that
floating-point calculations preserve every ground state.
Supplementary Note~8 gives
the numerical checks, pseudocode, and computational cost;
Supplementary Note~7 reports
separate tests of the released package.

\subsection{Instances and comparison design}
\label{sec:methods-design}

The quadratic benchmark uses random graphs with
$N\in\{100,200,500,1000\}$ spins, each with
$\Delta\in\{3,5\}$ neighbours. Couplings are drawn from a
Gaussian distribution, and there are no local fields.
The higher-order benchmark uses Gaussian spin glasses on regular
hypergraphs, planted 3-XORSAT with three constraints per variable,
and low-autocorrelation binary sequence (LABS) problems with dense
interactions.

We pass both Hamiltonians directly to OpenJij \texttt{sample\_hubo},
without converting higher-order terms to quadratic form. We use
$1000$ sweeps per returned sample.
Each original attempt receives the full time budget $T$;
each reduced attempt receives $T-t_{\mathrm{prep}}$ of
solver time. Thus, every reduced attempt includes the full
preprocessing cost, rather than sharing it across attempts.
In the development tests, library defaults can produce
different annealing schedules for the two formulations.
Separate checks and the confirmation test instead use the same
schedule for both formulations. Supplementary
Notes~3.8 and~4.4
give the instance counts, budgets, attempt schedules, reference
calculations, and recorded software versions; versions missing from
the original run records are identified there.

We calculate reference minima by tensor-network contraction in
double-precision arithmetic and independently evaluate the energies
of the returned configurations. These checks verify the minima
numerically; they do not provide exact-arithmetic certificates.
Planted 3-XORSAT minima are checked using arithmetic modulo
two, $\mathrm{GF}(2)$~\cite{kowalsky2022xorsat};
LABS uses published exact minimum energies
\cite{packebusch2016low}.
Larger spin-glass instances use best-known reference energies.
The benchmark runs themselves contribute to these references,
so the comparisons measure performance relative to the best
energies found, not against targets fixed independently before
testing. Certified and best-known references are never mixed
within the same benchmark setting.

The variable-fixing comparison uses $60$ instances generated
with new seeds. We compare equal total sample counts and
equal time budgets of $20$ and $60$~s. The two timed
comparisons reuse the same sequences of random seeds and are
therefore not independent repetitions. The primary measure compares
the best energies found with and without preprocessing on the same
instance (Eq.~\eqref{eq:paired-energy-difference}); it requires no
reference minimum. Best-known references fixed before these runs
are used only to compare how often each method reaches its target.
We validate the output of roof duality; it fixes no spins
on these instances and leaves the Hamiltonian unchanged.
Supplementary Notes~3.9 and~3.6 give the full SPVAR procedure
with multiple starts; Supplementary Note~3.7 and Table~7 give the
separate check on $100$ instances.

For the confirmation test, we fix and record the protocol
and analysis before generating $40$ new quadratic and
$40$ new three-body instances. Each instance is tested
with $50$ attempts per formulation. Each pair uses the same attempt
seed, and both formulations use one annealing schedule derived from
the original Hamiltonian. The elimination settings are fixed in advance.
Two tensor-network calculations, with independently chosen orders for
combining the energy tables, must give the same minimum. We also
check the returned configurations by evaluating their energies
independently in the original Hamiltonian. One quadratic instance exceeds
the time limit during its second verification calculation.
It is excluded from analyses that require a verified
minimum, but all $40$ quadratic instances remain in the
energy comparison
(Supplementary Note~6.1).

To test which cubic Max-Cut graphs fit the spin budget, we apply four
elimination orders to $100$ graphs at each even size $N=18,\ldots,40$.
Separate software tests of the AOC update rule and a greedy
solver use $20$ graphs per size and $200$ restarts per
formulation. After the removed spins are recovered, every
cut is evaluated in the original graph. These tests do
not measure physical AOC performance. Supplementary
Note~9 gives the update rule,
sign convention, settings, graph samples, and restart
analysis.

\subsection{Success measures and statistical analysis}
\label{sec:methods-statistics}

An attempt counts as successful when its best energy satisfies
$E_{\mathrm{best}}\leq E_{\mathrm{ref}}+10^{-4}$.
For each formulation in a given setting, let $p_i$ be the
observed fraction of successful attempts on instance $i$.
Averaging over the $M$ instances gives the pooled success
fraction, $\bar p_{\mathrm{succ}}=M^{-1}\sum_{i=1}^{M}p_i$.
For $0<\bar p_{\mathrm{succ}}<1$, we use this average to calculate
the time-to-solution index for $90\%$ success:
\begin{equation}
\label{eq:tts90}
\TTSindex
=
T\,
\frac{\ln(0.1)}{\ln(1-\bar p_{\mathrm{succ}})}.
\end{equation}
Equation~\eqref{eq:tts90} inserts the mean success fraction into the
restart formula and allows a noninteger number of attempts. The index can therefore be shorter than one
attempt. It is neither the mean nor the median of the
times to solution for individual instances.

To estimate success after a whole number of attempts,
we first calculate the chance of at least one success
for each instance and then average. Assuming independent
attempts with an unchanged success probability on each
instance, for any integer $n\geq1$ this gives
\begin{equation}
\label{eq:pooled-jensen}
\bar P(n)
=
\frac{1}{M}\sum_{i=1}^{M}\left[1-(1-p_i)^n\right]
\leq
1-(1-\bar p_{\mathrm{succ}})^n.
\end{equation}
The function $1-(1-p)^n$ is concave in $p$, which gives the
inequality. Applying the restart formula to the mean success
fraction can therefore overestimate average success when some instances
are harder than others. Rounding the pooled index to whole
attempts does not remove this limitation.

When every recorded attempt succeeds
($\bar p_{\mathrm{succ}}=1$), plots show one full budget
$T$. This is a display convention, not a confidence bound.
When no attempt succeeds, we infer no finite index;
observing no successes does not establish that the true
success probability is zero.

For the development tests, we calculate confidence intervals by
resampling instances with replacement while keeping the original
and reduced results together. This is paired bootstrap resampling.
Figure captions specify the intervals or error bars shown.

For the confirmation test, we subtract the original success fraction
from the reduced success fraction on each instance, then average
these differences. We calculate a two-sided $95\%$ confidence
interval from $20\,000$ bootstrap resamples. In each resample,
we first select instances and then select paired attempts within
each chosen instance. Confirmation requires the lower interval
bound to exceed zero. For the energy comparison, let $E^{\mathrm{R}}_{ir}$
and $E^{\mathrm{O}}_{ir}$ be the energies returned by paired attempt $r$
on instance $i$ for the reduced and original formulations, each the lowest
energy among that attempt's samples, evaluated in the original
Hamiltonian. The instance statistic and the reported value are
\begin{equation}
\label{eq:heldout-energy}
d_i=\operatorname{median}_{r}\left(E^{\mathrm{R}}_{ir}-E^{\mathrm{O}}_{ir}\right),
\qquad
\Delta E=\operatorname{median}_{i}\, d_i ,
\end{equation}
with the $95\%$ interval from the same $20\,000$-resample bootstrap over
instances and then paired attempts, recomputing both medians in each
resample. Instances count as wins, ties, or losses according to the sign
of $d_i$ with a tie tolerance of $10^{-4}$, and a two-sided sign test on
the non-tied instances compares the directions. Each instance has equal
weight; repeated attempts on one instance are not treated as independent
observations. This comparison needs no reference minimum. We also estimate
success after repeated attempts using
Eq.~\eqref{eq:pooled-jensen}. The pooled index is a summary,
not the primary test of improvement. Jeffreys smoothing
is used only to check how limited sampling affects the
restart estimates. Supplementary
Note~6.1 gives the complete analysis.

\section*{Code and data availability}

The core compiler, \texttt{walsh-elimination} version~0.1.0, is packaged
under the MIT licence with verification tests and a worked example; the
Zenodo identifier for the release is
\url{https://doi.org/10.5281/zenodo.22541332}.

\section*{Acknowledgements}
The author acknowledges support from the HORIZON EIC Pathfinder Challenges
project HEISINGBERG (grant 101114978), the EPSRC UK Multidisciplinary Centre
for Neuromorphic Computing (grant UKRI982) and the Weizmann--UK Make
Connection grant (grant 142568). The author used Anthropic's Claude and
OpenAI's ChatGPT for language editing, structural revision, and internal
consistency checks. The author independently verified all derivations,
numerical results, and conclusions and takes full responsibility for the
manuscript. The author thanks Stefan Boettcher for correspondence and for
drawing attention to his prior work on sparse spin-glass reductions.

\section*{Competing interests}
The author declares no competing interests.

\FloatBarrier
\bibliographystyle{unsrt}
\bibliography{references}

\end{document}